\documentclass[prl,twocolumn,showpacs,preprintnumbers,amsmath,%
amssymb,superscriptaddress,floatfix]{revtex4}

\usepackage{graphicx}
\usepackage{dcolumn}
\usepackage{bm}

\begin{document}

\title{Analytical approach to soliton ratchets in asymmetric potentials}

\author{Niurka R.\ Quintero}
\email{niurka@euler.us.es}
\affiliation{Departamento de F\'\i sica Aplicada I, E.\ U.\ P.,
Universidad de Sevilla, Virgen de \'Africa 7, 41011 Sevilla,
Spain}
\affiliation{Instituto Carlos I de F\'{\i}sica Te\'orica y
Computacional, Universidad de Granada, 18071 Granada, Spain}

\author{Bernardo S\'anchez-Rey}
\email{bernardo@us.es}
\affiliation{Departamento de F\'\i sica Aplicada I, E.\ U.\ P.,
Universidad de Sevilla, Virgen de \'Africa 7, 41011 Sevilla,
Spain}

\author{Mario Salerno}
\email{salerno@sa.infn.it} \affiliation{Dipartimento di Fisica
``E.R.Caianiello"
and Istituto Nazionale di Fisica della Materia (INFM),\\
Universit\'a di Salerno, via S. Allende, I84081 Baronissi (SA),
Italy;}


\date{\today}

\begin{abstract}
We use soliton perturbation theory and collective coordinate
ansatz to investigate the mechanism of soliton ratchets in a
driven and damped  asymmetric double sine-Gordon equation. We show
that, at the second order of the perturbation scheme, the soliton
internal vibrations can couple {\it effectively},
in presence of damping, to the
motion of the center of mass, giving rise to transport.  An
analytical expression for the mean velocity of the soliton is
derived. The results of our analysis confirm the internal mode
mechanism of soliton ratchets proposed in [Phys. Rev. E {\bf 65}
025602(R) (2002)].
\end{abstract}


\maketitle

\section{I. Introduction}

During the past years a great deal of attention has been devoted to the ratchet
effects both for point particles \cite{mag,han,cil,apph} and for extended
systems \cite{mar,sav,fal}. Some experimental realization of these models can
be found in \cite{optical,lee,fal,carapella} (see \cite{rei} for a recent
review). Ratchet-like systems such as systems of two particle with internal
degree of freedom \cite{cil}, periodic rocket ratchets (deterministic and
stochastic) \cite{bar} and temperature ratchets \cite{rei1}, have also been
considered. For infinite dimensional systems described by nonlinear partial
differential equations (PDE) of soliton type, ratchet phenomena were
investigated  both in the case of asymmetric potentials in presence of
symmetric forces \cite{mn}, and in the case of symmetric potentials with
asymmetric forces \cite{sz,flach}. In both cases the ratchet phenomenon
manifests as an unidirectional motion of the soliton, similar to the drift
motion occurring for point particle ratchets (from here the name of soliton
ratchets). A symmetry approach to the phenomenon, which allows to establish
conditions for the occurrence of soliton ratchets, was developed in Ref. \cite
{flach}.  This approach, although useful for predicting the phenomenon, does
not provide information about the actual mechanism responsible for the
unidirectional motion. The mechanism underlying soliton ratchets was recently
proposed by two of us for the case of a perturbed asymmetric double sine-Gordon
equation driven by symmetric forces \cite{mn}, and extended in Ref. \cite{sz}
to the case of a perturbed sine Gordon system in presence of asymmetric
drivers. In both cases the phenomenon was ascribed to the existence of an
internal mode mechanism which, in presence of damping, is able to couple
soliton internal vibrations to the translational mode, producing in this way
transport. The phenomenon was described as follows: the energy pumped by the ac
force into the soliton internal mode is converted into net dc motion by the
coupling of the internal vibration with the center of mass induced by the
damping. This internal mode mechanism was confirmed by numerical and analytical
investigations for the case of the sine Gordon system with asymmetric periodic
forces \cite{sz,flach,luis} and by direct numerical investigations for the case
of soliton ratchets of the asymmetric double sine-Gordon equation \cite{mn}. In
the last case, a detailed numerical investigation (see \cite{mn}) showed the
following facts:

{\bf (i)} The presence of damping and the asymmetry of the
potential are both crucial ingredients for the existence of the
net motion of the soliton (in presence of the ac force, but in
absence of damping, the asymmetry of the potential does not
produce transport). {\bf (ii)} The effect of the damping is to
couple the internal vibrations  of the kink  to the motion of the
center of mass. {\bf (iii)} For fixed values of the amplitude and
frequency of the ac force there is an optimal value of the damping
for which the transport is maximal. {\bf (iv)} The direction of
the motion is fixed by the asymmetry of the potential and is
independent on initial conditions.  {\bf (v)} For fixed values of
the damping the average velocity of the kink shows a resonant
behavior as a function of the frequency of the ac force . {\bf
(vi)} At low damping and higher forcing strengths, current
reversals in the kink dynamics can occur. This phenomena was
ascribed in Ref. \cite{mn} to the soliton-phonon interaction
rather than to the internal mode mechanism.

These  points were found to be valid also for different  soliton
ratchet systems \cite{sz,cos}. In Ref. \cite{cos}, the existence
of an optimal value of the damping constant  which maximizes the
transport, was ascribed to the relativistic nature of the kink
dynamics and the possibility of nonzero currents in absence of
damping, was also found. Since most of the analysis of Ref. \cite{mn} has been based
on numerical simulations, it is of interest to provide an
analytical confirmation of the above results.

The aim of this paper is to present an analytical investigation of
soliton ratchets of the asymmetric double sine-Gordon equation
which confirms the aforementioned internal mode mechanism proposed
in Ref. \cite{mn} as well as points {\bf (i)} - {\bf (vi)} listed
above. To this end, we use perturbation theory and a collective
coordinate ansatz for the soliton shape to derive ordinary
differential equations (ode) for the center of mass and for the
soliton width. We show that, to second order of perturbation
theory, soliton width vibrations couple {\it effectively} to the
motion of the center of mass via the damping in the system. By
using the collective coordinate equations we are able to derive an
analytical expression of the mean drift velocity of the kink as a
function of the system parameters. The analysis is shown to be in
good agreement with numerical simulation and with the results {\bf
(i)} - {\bf (vi)} obtained in Ref. \cite{mn}.

The paper is organized as follow: in  section II we introduce our
model and discuss its main properties. In section III we derive
the dynamical equations for soliton collective coordinates and
obtain an  expression of the average kink velocity as a function
of the system parameters. In section IV we compare our analytical
results with direct numerical simulations and discuss them in
connection with previous work. Finally, in section V the main
conclusions of the paper are summarized.

\section{II. The Model}

Let us  consider the following perturbed asymmetric double
sine-Gordon equation (ADSGE)
\begin{eqnarray}
\phi _{tt}-\phi _{xx} &+&\sin (\phi )+\lambda \cos (2\phi )=
F(x,t,\phi,\phi_{t},...) \equiv \nonumber
\\
f(t) &-&\alpha \phi _{t},  \label{eq1}
\end{eqnarray}
where $\lambda \in [-1,1]$ is a parameter related to the asymmetry of the
nonlinear Klein-Gordon potential, $\alpha$ is a damping constant and
$f(t)=\epsilon \sin (\omega t+\theta_{0})$ is a periodic force with amplitude
$\epsilon$, frequency $\omega $ and phase $\theta _{0}$. This system is
connected with interesting physical problems such as arrays of inductively
coupled asymmetric SQUIDs of the type considered in Ref. \cite{zapata}. A
mechanical analogue of Eq. (\ref{eq1}) in terms of  a chain of double pendulum
was given in Ref. \cite{ms85}. For $F=0$, Eq. (\ref{eq1}) has an hamiltonian
structure with Hamiltonian (energy)
\begin{eqnarray}
\label{eqhami} H & = & \int_{-\infty}^{+\infty} dx \left\{
\frac{1}{2} (\phi_t^{2} + \phi_{x}^{2}) + U(\phi) \right\},
\end{eqnarray}
and momentum
\begin{eqnarray} \label{eqhami1}
P & = & - \int_{-\infty}^{+\infty} dx \phi_{x} \phi_t.
\end{eqnarray}
The potential in Eq.~(\ref{eqhami}) is $U(\phi)=C-\cos(\phi)+(\lambda/2) \sin(2
\phi)$ and $C=\cos(\phi_{0}) - (\lambda/2) \sin(2 \phi_{0})$ (notice that for
$\lambda \ne 0$ the potential is asymmetric in the field variable). Similar to
the sine-Gordon equation, the energy and the momentum are both conserved
quantities for the unperturbed ADSGE. In this case one can show \cite{mn} that
Eq. (\ref{eq1}) has an exact kink (anti-kink) solution of the form
\begin{equation}
\label{eq2} \phi^{\pm}_{K}=\phi_{0}+2\tan^{-1}
\left\{\frac{{\rm sign}(\lambda)\,A\,B}{A-1- \eta \sinh \left[\pm
\frac{\xi}{2}\sqrt{\frac{AB}{|\lambda |}}\right]}\right\}
\end{equation}
where $\xi=(x - V t)/\sqrt{1-V^{2}}$, $A=\sqrt{1 + 8
\lambda^{2}}$, $\eta=2\lambda\sqrt{2(1+A)}$, $B=\sqrt{2(4\lambda
^{2}-1+A)}$ and $\phi_{0} = \arcsin[(1-A)/(4 \lambda)] + 2 \pi n$
(the plus and minus signs refer to the kink and antikink
solutions, respectively). In the limit $\lambda \rightarrow 0$
(zero asymmetry), Eq. (\ref{eq2}) reduces to the well known
soliton solution of sine-Gordon equation (SGE).

\section{III. Collective Coordinate Analysis}

The term $F(x,t,\phi,\phi_{t})$ added to the ADSGE can be
considered as a small perturbation acting on the system. In this
case the energy and the momentum will depend on time so that it is
natural to assume an {\it ansatz} for the perturbed kink of the
form
\begin{eqnarray}
\label{xl} \phi &=& \phi_{0}+2\tan^{-1} \left\{\frac{{\rm
sign}(\lambda)\,A\,B}{A-1- \eta \sinh \left[ \frac{x-X(t)}{W(t)}
\right]}\right\},
\end{eqnarray}
where $X(t)$ and $W(t)$ represent dynamical collective coordinates
(CC) corresponding to the center of mass and the width of the
kink, respectively (similar approach was introduced in \cite{ms}
for the SGE). In the following we consider only kink solutions
since the analysis of anti-kinks will follow from it without
difficulty. In absence of perturbation the kink moves with
constant velocity $V$ and constant width $W(t)=W_{s}=W_{0}
\sqrt{1-V^{2}}$ ( $W_{0} = 2 \sqrt{|\lambda|/[A B]}$ ), so that
$X(t)= V t$. Notice that {\it ansatz} (\ref{xl}) is consistent
with the linear stability analysis  performed in Ref. \cite{mn},
showing the existence of an internal mode frequency $\Omega_{I}$
below the phonon band (this mode disappears in the sine-Gordon
limit $\lambda = 0$). From the spatial profile of the
corresponding localized eigenfunction, one can see  that the
internal mode is linked to vibrations of the kink  width, this
suggesting the choice of {\it ansatz} (\ref{xl}).

By substituting Eq. (\ref{xl}) into Eqs. (\ref{eqhami}),
(\ref{eqhami1}), and differentiating with respect to time, one
obtain, after straightforward calculations (for details see
\cite{gtwa}), that $X(t)$, $P(t)$ and $W(t)$ satisfy the following
system  of nonlinear ode
\begin{subequations}
\begin{eqnarray}
\label{cc1} & \, & \frac{dX}{dt} = \frac{P(t) W(t)}{R^{2}
I_{1}} - \frac{I_{2}}{I_{1}} \dot{W}, \\
\nonumber & \, & \dot{W}^{2}-2 W \ddot{W} - 2 \alpha W \dot{W} = -
\frac{I_{1}}{K} + \frac{W^{2}}{K} \left[ \frac{P^{2}}{R^{4} I_{1}}
+ \right.
\\  & \, & \left.
 \left({\rm sign}(\lambda) I_{4} -
q \frac{I_{2}}{R I_{1}} \right) \frac{2 f(t)}{R} + \frac{2}{R^{2}}
g \right],\label{cc3}
\end{eqnarray}
\end{subequations}
where the momentum $P$ is a solution of the equation
\begin{equation}
\label{cc2}
\frac{dP}{dt} = -\alpha P - q f(t),
\end{equation}
and
\begin{eqnarray}
\hspace{-0.5cm} \hspace{-0.5cm} & \, & I_{i}=
\int_{-\infty}^{+\infty} \frac{\cosh^{2}(\theta) \theta^{i-1} d
\theta}
{[(A-1-\eta \sinh[\theta])^{2}+A^{2} B^{2}]^{2}}, \, i=1,2,3; \nonumber \\
\hspace{-0.5cm} & \, & I_{4}= \int_{-\infty}^{+\infty}
\frac{\cosh(\theta) \theta d \theta} {(A-1-\eta
\sinh[\theta])^{2}+A^{2} B^{2}}, \nonumber \\ & \, & K=
\left(I_{3} - \frac{I_{2}^{2}} {I_{1}}\right); \hspace{.3cm} R=2 A
B \eta, \hspace{.3cm} g = \frac{I_{1}(\lambda) A^{3} B^{3}
\eta^{2}}{2 |\lambda|}. \nonumber
\end{eqnarray}
Notice that the above integrals depend on the asymmetry parameter
$\lambda$ and are finite for any value of $\lambda$ (they can be
easily calculated by numerical tools). In Ref. \cite{luis} a
similar  approach was used to show that the internal mode
mechanism give rise to net motion in the SG equation perturbed by
two harmonic forces.

In order to solve the system of equations (\ref{cc1})-(\ref{cc3})
and (\ref{cc2}) we use the following initial conditions:
$dX(0)/dt=0$, $P(0)=0$ and $W(0)=W_{0}$. We solve first the
non-coupled linear equation for the momentum. From Eq. (\ref{cc2})
we obtain
\begin{eqnarray}
\label{soluP} & \, & P(t)  = \frac{- q
\epsilon}{\sqrt{\alpha^{2}+\omega^{2}}} \sin[\omega t + \theta_{0}
+ \delta] +  \nonumber\\ & \, & {\rm e}^{-\alpha t} [P(0) +
\frac{q \epsilon}{\sqrt{\alpha^{2}+\omega^{2}}} \sin(\theta_{0} +
\delta)],
\end{eqnarray}
where $\tan (\delta) = -\omega/\alpha$ and the last term in the
r.h.s, being a transient, will be neglected (we are interested in
the stationary regime  $t >> \alpha^{-1}$). Notice from Eq.
(\ref{cc3}) that the width of the kink is affected directly by the
ac force with a frequency $\omega$ and indirectly by the
translational motion (the momentum) with a frequency $2 \omega$
(see Ref. \cite{gtwa}b). Since it is difficult to find an exact
solution of this equation, we will search for an approximate
solution for $W(t)$ in the form of a power series in $\epsilon$
\begin{equation}
\label{expansion} W(t) =  W_{0} + \epsilon W_{1}(t) + \epsilon^{2}
W_{2}(t) + O(\epsilon^{3}).
\end{equation}
Inserting (\ref{expansion}) into (\ref{cc3}) and taking terms of
the same order in $\epsilon$, we obtain,  after a transient time
($t
>> \alpha^{-1}$), the following set of linear equations.
At  order $O(\epsilon)$  we have
\begin{equation}
\ddot{W}_{1} + \Omega^{2} W_{1} + \alpha \dot{W}_{1} = W_{0} G (\lambda)
\sin(\omega t + \theta_{0}), \label{eql1}
\end{equation}
where
\begin{eqnarray}
\label{omega}
\Omega^{2} & = & \frac{2  g(\lambda)}{R^{2} k(\lambda)},
 \\
G(\lambda) & = & \frac{q I_{2}-{\rm sign}(\lambda) I_{1} I_{4} R
}{(I_{1} I_{3} -I_{2}^{2}) R^{2}},
\end{eqnarray}
while at the second order ($O(\epsilon^{2})$) we obtain
\begin{eqnarray} \label{eql2}
& \, & \ddot{W}_{2} + \Omega^{2} W_{2} + \alpha \dot{W}_{2} =
\frac{\dot{W}^{2}_{1}}{2 W_{0}} + \frac{\Omega^{2} W_{1}^{2}} {2
W_{0}} -  \\ \nonumber & \, &  \frac{W_{0} q^{2} \sin^{2}[\omega t
+ \theta_{0} + \delta]} {2 R^{4}(I_{1} I_{3} - I_{2}^{2})
(\alpha^{2} + \omega^{2})} + G (\lambda) W_{1} \sin(\omega t +
\theta_{0}).
\end{eqnarray}
Eqs.\ (\ref{eql1}) and (\ref{eql2}) correspond to linear, damped
and driven oscillators with characteristic frequency $\Omega$. It
can be shown that for $\lambda \in [-1,1]$ the values of $\Omega$
are quite close to the internal mode frequency $\Omega_{I}$ of the
ADSGE computed in \cite{mn} (the maximum difference between them
being $\approx 0.1$). Eqs.\ (\ref{eql1}) and (\ref{eql2}) can be
solved exactly. The homogeneous part of the solutions of both
equations decays to zero, after a transient time, if $\alpha^{2} -
4 \Omega^{2} < 0$. The particular solutions of (\ref{eql1}) and
 (\ref{eql2}) are given, respectively, by
\begin{eqnarray}
\label{soll1} \hspace{-0.4cm} & \, & W_{1}(t) =  - \frac{W_{0}
G(\lambda) \cos(\omega t + \theta_{0} +
\tilde{\theta})}{\sqrt{(\Omega^{2} - \omega^{2})^{2} +
\alpha^{2} \omega^{2}}}, \\
\label{soll2}\hspace{-0.4cm} & \, & W_{2}(t) = \frac{A_{0}}
{\Omega^{2} [(\Omega^{2}-\omega^{2})^{2}+ \alpha^{2} \omega^{2}]}
+ \\ \nonumber\hspace{-0.4cm} &\,& \frac{A_{1} \sin(2 \omega t + 2
\theta_{0} + 2 \tilde{\theta} + \tilde{\theta}_{1})}
{[(\Omega^{2}-\omega^{2})^{2}+ \alpha^{2} \omega^{2}]
\sqrt{(\Omega^{2}-4 \omega^{2})^{2}+ 4 \alpha^{2} \omega^{2}}} +
\\ \nonumber\hspace{-0.4cm} & \, & \frac{A_{2} \sin(2 \omega t + 2
\theta_{0} + 2 \delta + \tilde{\theta}_{1}) + A_{3} \cos(2 \omega
t + 2 \theta_{0} + \tilde{\theta} + \tilde{\theta}_{1})}
{[(\Omega^{2}-\omega^{2})^{2}+ \alpha^{2} \omega^{2}]
\sqrt{(\Omega^{2}-4 \omega^{2})^{2}+ 4 \alpha^{2} \omega^{2}}},
\end{eqnarray}
where $\tan(\tilde{\theta})=(\Omega^{2} - \omega^{2})/(\alpha
\omega)$, $\tan(\tilde{\theta}_{1})=(\Omega^{2} - 4 \omega^{2})/(2
\alpha \omega)$ and
\begin{eqnarray}
\label{coeff} & \, & A_{0} = A_{1} - A_{2} + \frac{W_{0} G^{2}
\Omega^{2}}{2}, \nonumber
\\ \nonumber
& \, & A_{1} = W_{0} G^{2}(\lambda) \left\{\frac{\Omega^{2}}{4}
-\frac{\omega^{2}}{4} \right\},
\\
\nonumber & \, & A_{2} = \frac{W_{0} q^{2}
[(\Omega^{2}-\omega^{2})^{2}+ \alpha^{2} \omega^{2}]}{4 R^{4}
(I_{1} I_{3} - I_{2}^{2}) (\alpha^{2}+\omega^{2})}, \\
\nonumber & \, & A_{3} = \frac{W_{0} G^{2}(\lambda)
\sqrt{(\Omega^{2}-\omega^{2})^{2} +\omega^{2} \alpha^{2}}}{2}.
\end{eqnarray}
Substituting Eqs.\ (\ref{soluP}) and (\ref{expansion}) into Eq.\
(\ref{cc1}), we obtain that the velocity of the kink up to the
second order in $\epsilon$ is given by
\begin{eqnarray}
\nonumber \frac{dX}{dt} &=& \epsilon \left[ \frac{ -q W_{0}
\sin(\omega t+ \theta_{0} + \delta)} {R^{2} I_{1}(\lambda)
\sqrt{\alpha^{2}+ \omega^{2}}} -
\frac{I_{2}(\lambda)}{I_{1}(\lambda)} \dot{W}_{1} \right] +
\\ \label{velo}
& \, & \epsilon^{2} \left[ \frac{ -q W_{1} \sin(\omega t+
\theta_{0} + \delta)} {R^{2} I_{1}(\lambda) \sqrt{\alpha^{2}+
\omega^{2}}} - \frac{I_{2}(\lambda)}{I_{1}(\lambda)} \dot{W}_{2}
\right].
\end{eqnarray}
By taking the average value of this velocity over one period
($T=2\pi/\omega$), $\langle dX/dt \rangle \equiv \langle V \rangle
= (1/T) \int_{0}^{T}  (dX/dt) d\tau $, we finally obtain
\begin{eqnarray}
\hspace{-0.6cm} \langle V \rangle & = & \frac{- \epsilon^{2} q
W_{0} G(\lambda) \Omega^{2} \alpha}{2 R^{2} I_{1}(\lambda)
(\alpha^{2}+ \omega^{2}) [(\Omega^{2} - \omega^{2})^{2} +
\alpha^{2} \omega^{2}]}. \label{solx2a}
\end{eqnarray}

Eqs. (\ref{velo}) and (\ref{solx2a}) represent the main result of the
paper. From their analysis the following important conclusions can
be made. First, we notice that the non-zero average velocity is
due to the {\it effective interaction between the translational}
($P(t)$) {\it and internal mode} ($W_{1}(t)$) represented by the
first term in the second bracket of (\ref{velo}). Second, Eq.
(\ref{solx2a}) shows that the average velocity does not depend on
the initial phase. Indeed, the direction of the motion is
determined only by $G(\lambda)$, i.e. by the parameter $\lambda$
which controls the asymmetry of the potential. It is not difficult
to check that $G(\lambda) > 0$ ($G(\lambda) < 0$) for $\lambda >
0$ ($\lambda < 0$) and that $G(\lambda)$ vanishes at $\lambda=0$.
Third, from (\ref{solx2a}) we also see that for a given frequency
there is an optimal value of the damping to achieve the maximal
mean velocity (see Fig.\ 4 of \cite{mn}). This optimal value can
be easily calculated as a function of $\omega$ and $\lambda$ (see
the solid line of Fig.\ \ref{fig-alphaopt}), and is given by
\begin{equation}
\alpha_{opt} = \sqrt{ \frac{-D- \omega^{4}+\sqrt{(D+\omega^{4})^2+12 D
\omega^{4}}}{6 \omega^{2}} }, \label{alphaopt}
\end{equation}
where $D=(\Omega^{2}-\omega^{2})^2$.

Fourth, the activation of the internal mode alone, without any phonons present
in the system, does not allow the damping to rectify the motion. By varying
$\alpha$ in Eq.\ (\ref{solx2a}) we can not change the sign of the average
velocity. This feature confirms the prediction of \cite{mn}, where the
rectification of the movement for small damping and large amplitude of the ac
force was related to the excitation of phonons for a given choice of
parameters.

Moreover, the existence of a resonant behavior of the mean
velocity as a function of the frequency $\omega$ is also
qualitatively confirmed by Eq. (\ref{solx2a}).  As reported in
Ref. \cite{mn} (see Fig.\ 6 of this paper), $\langle V \rangle$
becomes maximum when $\omega$ approaches the internal mode
frequency $\Omega_{I} \approx \Omega$. This agreement, however, is
only qualitative since the resonant peak is very close to the edge
of the phonon band so that phonons are easily excited in the
system. In fact, Eq. (\ref{soll1}) together with Eq. (\ref{soll2})
show that the kink oscillates with two frequencies $\omega$ and
$2\omega$. Close to the resonance $2\omega \approx 2\Omega$ one is
inside the phonon band so that phonons get excited and the CC
analysis becomes unaccurate.

Notice that Eq.\ (\ref{solx2a}) is valid for small $\epsilon$ and
for times ($t>> \alpha^{-1}$), thus for the zero-damping case
nothing can be inferred  about the mean velocity. It is of
interest to investigate the zero damping case separately so to
check the role played by the damping in the phenomenon. For
$\alpha=0$ the momentum equation (\ref{cc2}) simplifies as well as
the collective coordinate equations (\ref{cc1})-(\ref{cc3}). From
Eq. (\ref{cc2}) (with $\alpha=0$) we have that a solution
satisfying the initial condition $P(0)=0$ is readily obtained as
\begin{equation}  \label{eqP0}
P(t) = \frac{q\epsilon}{\omega} [\cos (\omega t+\theta_0)-\cos(\theta_0)].
\end{equation}
By substituting this equation and Eq. (\ref{expansion}) into Eq.
(\ref{cc1}) we obtain that the kink velocity at order $\epsilon$
is
\begin{equation}  \label{eqX0}
\frac{dX}{dt}= \frac{q W_{0} \epsilon}{R^2 \omega I_1(\lambda)}
[\cos(\omega t +
\theta_{0})-\cos\theta_0] 
-\frac{I_2(\lambda)}{I_1(\lambda)} \epsilon \dot{W}_1.
\end{equation}
The  first order correction to the soliton width, $W_1$, can be
calculated solving Eq.\ (\ref{eql1}) with $\alpha=0$. For $\omega
<< \Omega$ we obtain
\begin{eqnarray}
W_{1}(t)&=& - \frac{W_{0} G(\lambda) \cos \theta_0
\sqrt{\omega^2+\Omega^2 \tan^2{\theta_0}}}{\Omega (\Omega^{2} -
\omega^{2})} \sin(\Omega t + \varphi) \nonumber \\  & & +
\frac{W_0 G(\lambda)}{\Omega^{2} - \omega^{2}} \sin(\omega t +
\theta_{0}),
\end{eqnarray}
where $\tan{\varphi}=(\Omega/\omega) \tan{\theta_0}$ in order to
fulfill the initial conditions $W_1(0)=0$ and $\dot{W}_1(0)=0$. It
is clear that the velocity of the kink depends on the initial
phase, $\theta_0$ so that, by averaging  over the phase, one
obtains zero transport. A straightforward calculation shows that
the same result is true also at order $\epsilon^2$.
This implies that the ratchet effect in the zero-damping case does not exist.
As we will see in the next section, this conclusion is confirmed by numerical
simulations of Eq. (\ref{eq1}).

\section{III. Numerical simulations and discussion}

The CC analysis of the previous section neglects the presence of phonons in the
system and is based on a particular  {\it ansatz} for the soliton shape in Eq.
(\ref{xl}). Moreover, the approximated solution (\ref{solx2a}) is valid only
when the perturbations are small enough ($\epsilon<<1$). To check our results
we compare numerical solutions of Eqs. (\ref{cc1})-(\ref{cc3}) and the
approximated solution in Eq. (\ref{solx2a}), with direct numerical integrations
of Eq. (\ref{eq1}). Numerical simulations of (\ref{eq1}) were performed by
using a 4th order Runge-Kutta scheme \cite{numrec} with steps $\Delta t=0.01$,
$\Delta x=0.1$ in the finite length domain $x \in [-50,50]$, taking into
account $200$ time periods. In order to obtain the numerical solutions of the
CC equations we have integrated Eqs. (\ref{cc1})-(\ref{cc3}) with the routine
DIVPRK of the IMSL library \cite{imsl} which uses the Runge-Kutta-Verner sixth
order method.

\begin{figure}
\vspace*{2mm}
\includegraphics[width=6.5cm,height=8.0cm,angle=-90]{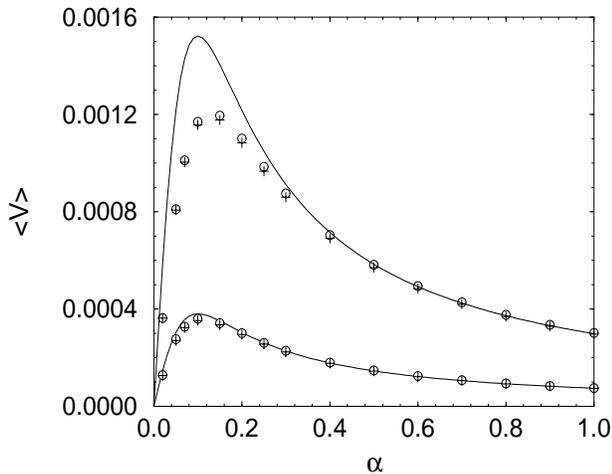}
\caption{\label{fig1} Dependence of the mean kink velocity on the
damping coefficient for two different values of strength of the
ac force $\epsilon=0.1$ (upper curves) and $\epsilon=0.05$ (lower
curves). Other parameters are fixed as $\lambda=-0.5$,
$\omega=0.1$, $\theta_{0}=\pi/2$.  The solid lines represent the
approximate solution of the CC theory (Eq.\ (\ref{solx2a}));
whereas circles and pluses refer, respectively, to PDE numerical
simulations of (\ref{eq1}) and numerical solutions of the CC
equations (\ref{cc1})-(\ref{cc3}).}
\end{figure}

\begin{figure}
\vspace*{2mm}
\includegraphics[width=6.5cm,height=8.0cm,angle=-90]{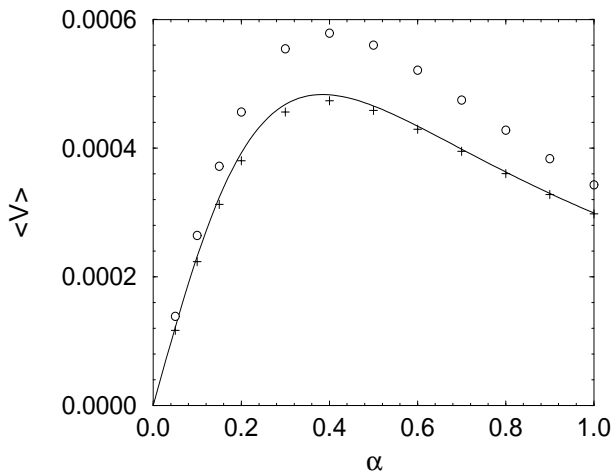}\\
\caption{\label{fig2} Mean kink velocity versus the damping
coefficient. Parameters are $\lambda=-0.5$, $\epsilon=0.1$,
$\omega=0.4$, $\theta_{0}=\pi/2$. The solid line represents the
approximated solution of the CC theory (Eq.\ (\ref{solx2a}));
whereas circles and pluses are the results of the numerical
simulations of the PDE (\ref{eq1}) and the numerical solutions of
the CC equations (\ref{cc1})-(\ref{cc3}), respectively. }
\end{figure}

Figures~\ref{fig1} and \ref{fig2} show the mean velocity dependence on the
damping coefficient for different values of the amplitude and frequency of the
ac force. From Fig.~\ref{fig1} we see that for low values of $\epsilon$ there
is a very good agreement between the CC results and PDE simulations in the
whole range of $\alpha$. However, when the amplitude of the ac force is
increased, the approximated solution (\ref{solx2a}) deviates from the numerical
solution of the CC equations and from the PDE simulations, mostly in the
optimal damping region.
In Fig.~\ref{fig2} we show the same dependence for an higher value of the
driver frequency ($\omega=0.4$). We see that although the approximated solution
(\ref{solx2a}) and the numerical solution of the CC equations are quite close,
there is a discrepancy with the PDE results starting from the peak and
extending to higher values of $\alpha$, this indicating that the CC {\it
ansatz} is valid only for $\omega << \omega_{ph}$ ($\omega_{ph}=1.2699$ for
$\lambda=-0.5$).

\begin{figure}
\vspace*{2mm}
\includegraphics[width=6.5cm,height=8.0cm,angle=-90]{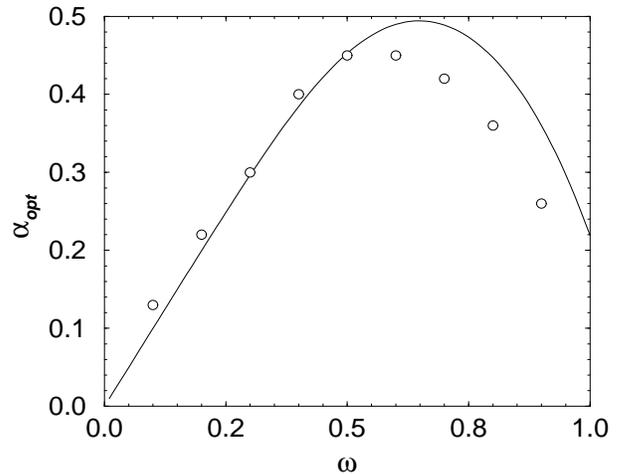}
\caption{\label{fig-alphaopt} Optimal value of the damping as a function of the
frequency of the ac force for $\lambda=-0.5$. The solid line represents the
results obtained from Eq.~(\ref{alphaopt}) while the circles refer to numerical
simulations of Eq.~(\ref{eq1}).}
\end{figure}

In Fig.~\ref{fig-alphaopt}, we check Eq.~(\ref{alphaopt}) for the  optimal
value $\alpha_{opt}(\omega)$ of the damping constant for $\lambda=-0.5$. We can
see that this value increase up to $\omega \approx 0.7$ and after that
$\alpha_{opt}(\omega)$ decreases. Again, a good agreement between the CC theory
and PDE results is found at small $\omega$ values.

\begin{figure}
\vspace*{2mm}
\includegraphics[width=6.5cm,height=8.0cm,angle=-90]{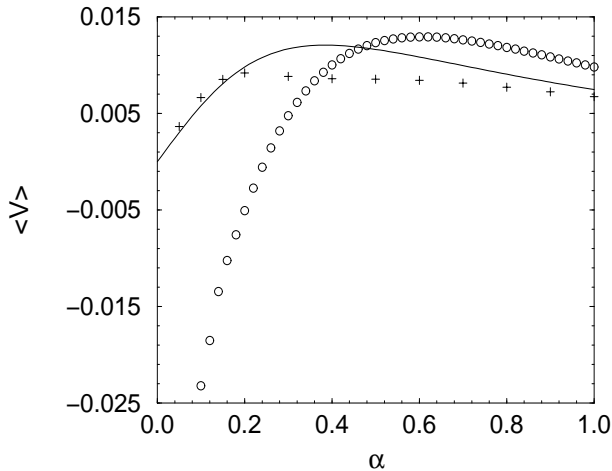}
\caption{\label{fig4-sal} Mean kink velocity versus the damping coefficient for
$\lambda=-0.5$; $\epsilon=0.5$, $\omega=0.4$, $\theta_{0}=\pi/2$. The solid
line represents the approximated solution of the CC theory (Eq.~(\ref{solx2a}))
while circles and pluses denote the results of the numerical simulations of the
PDE (\ref{eq1}) and of the numerical solutions of the CC equations
(\ref{cc1})-(\ref{cc3}), respectively. The PDE results show that for very small
values of damping ($\alpha<0.1$) pairs of kink and antikink appear. }
\end{figure}

\begin{figure}
\vspace*{2mm}
\includegraphics[width=6.5cm,height=8.0cm,angle=-90]{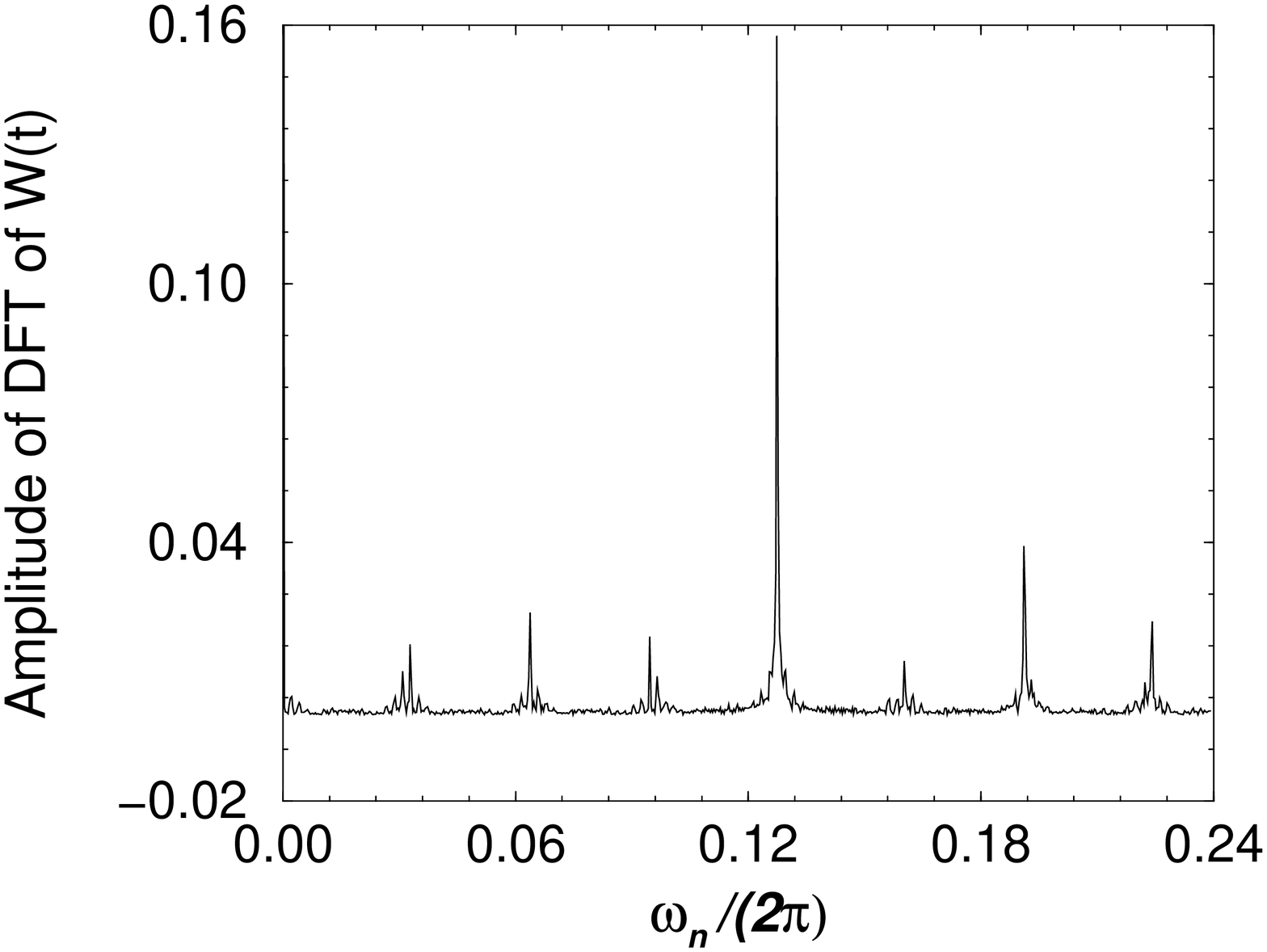} \\
\includegraphics[width=6.5cm,height=8.0cm,angle=-90]{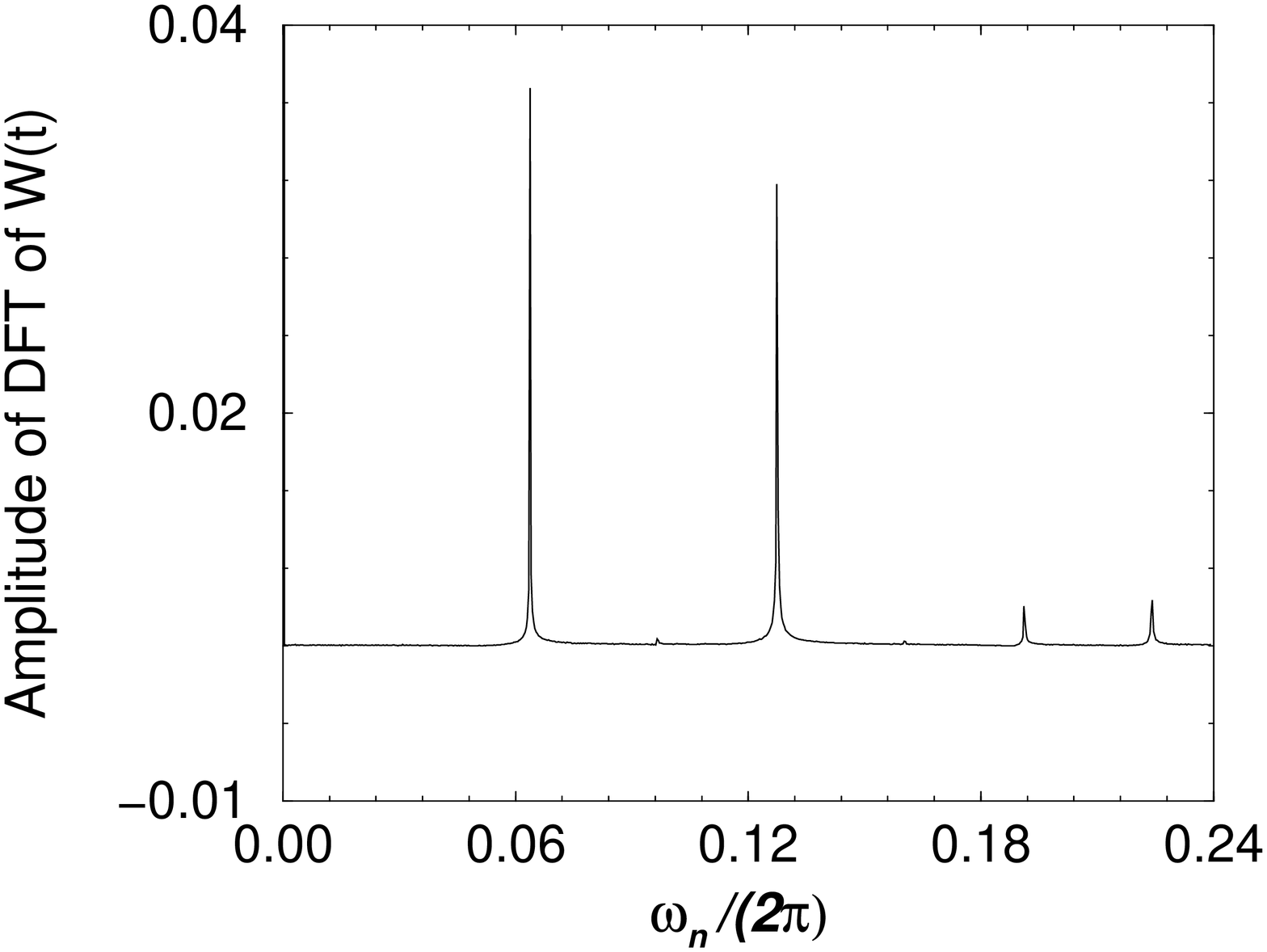}
\caption{\label{dft1} DFT of the width of the kink, $W(t)$,
obtained from the numerical solutions of the PDE for the same
parameters of Fig.\ \ref{fig4-sal} and $\alpha=0.1$ (upper panel)
and  $\alpha=0.7$ (lower panel, when the current is rectified).
The upper panel refers  to the case in which a current reversal
occours. The relevant frequencies in the spectrum are:
$\omega_{1}=2 \omega=0.8$; $\omega_{2}=3 \omega=1.2$; $\omega_{3}=
\omega=0.4$ and $\omega_{4}= (7/2) \omega= 1.4$ (this last being
inside the phonon band). The lower panel correspond to the case in
which the current is rectified. The main frequencies in this case
are located at $\omega_{1}=\omega=0.4$ and $\omega_{2}=2
\omega=0.8$ away from the phonon's band. The frequency of the
internal mode and the lower phonon's frequency are $\Omega_{I} =
1.056$ and $\omega_{ph}=1.269$, respectively.}
\end{figure}

By increasing the amplitude $\epsilon$ of the driver, reversal of current can
also  occur \cite{mn}. This is shown in Fig.~\ref{fig4-sal} from which we see
that as  $\alpha$ is increased the mean velocity computed from PDE simulations
displays a crossover from negative to positive values (i.e. a current reversal
occurred). This phenomenon is described neither by the CC analysis (notice that
Eq.~(\ref{solx2a}) predicts a positive average velocity for all $\lambda < 0$),
nor by the numerical solutions of the CC analysis depicted in the figure. This
agrees with the claim made in Ref.~\cite{mn} that reversal currents do not
depend on the internal mode mechanism, but rather on the existence of phonons
in the system. In order to show the presence of phonon modes when this
phenomenon appears, we have plotted  in Fig.~\ref{dft1} the discrete Fourier
Transform (DFT) of the kink's width $W(t)$ (obtained from the numerical
simulations of the PDE equation) for two values of $\alpha$, i.e. $\alpha=0.2$
(before crossover occurs) and $\alpha=0.7$ (inside the rectified motion
region). In the former case (upper panel), one of the frequencies of the
oscillations of $W(t)$ lies inside the phonon band, so phonons are clearly
excited. In the latter case (lower panel) the main frequencies of the spectrum
are well below the lower phonon edge ($w_{ph}/(2\pi)\approx 0.2$). From this we
conclude that phonon modes are important for current reversals and a theory
based on the internal mode alone (as the one  presented here) cannot describe
properly this phenomenon.

\begin{figure}
\vspace*{2mm}
\includegraphics[width=6.5cm,height=8.0cm,angle=-90]{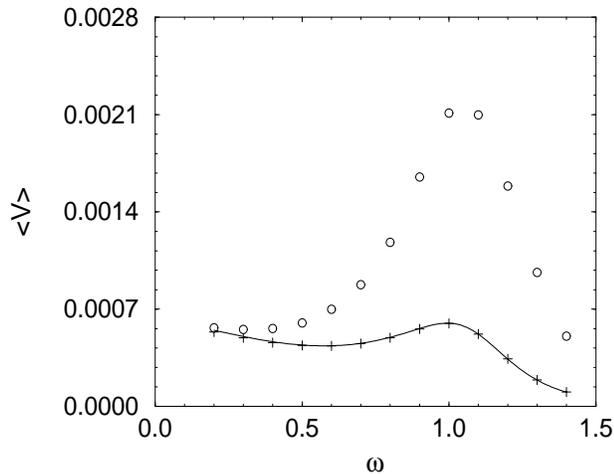}
\caption{\label{fig3-sal} Mean velocity of the kink as a function
of $\omega$. The pluses superimposing the solid line show a good
agreement between the approximated (solid line) and the numerical
(pluses) solutions of the CC theory. Results obtained from the
integration of the Eq.\ (\ref{eq1}) (circles) coincide with the CC
theory only for smaller values of $\omega$. Parameters are fixed
as  $\lambda=-0.5$, $\epsilon=0.1$, $\theta_{0}=\pi/2$ and
$\alpha=0.5$.}
\end{figure}

In Ref.~\cite{mn} a resonant behavior of mean velocity  as a function of
$\omega$ was also reported (see the peak of $\langle V \rangle$ at $\omega
\approx \Omega_{I}$ in Fig.\ 3 of this paper). This feature is also confirmed
by Eq.~(\ref{solx2a}) (notice that the denominator of this expression has a
minimum at $\omega=\Omega \approx \Omega_{I}$) although, in this case, the
agreement with PDE results is only qualitative. This is shown in Fig.\
\ref{fig3-sal} from which we see the presence of a resonant structure, with
good agreement at small frequencies and large deviations  from PDE results at
higher $\omega> 0.6$ values. The agreement at low frequencies can be understood
from the fact that phonons in this case are hardly excited and the CC
description becomes accurate. On the contrary, when $\omega$ gets close to the
resonant peak, phonons are easily excited  and the CC analysis becomes
unadequate.

\begin{figure}
\vspace*{2mm}
\includegraphics[width=6.5cm,height=8.0cm,angle=-90]{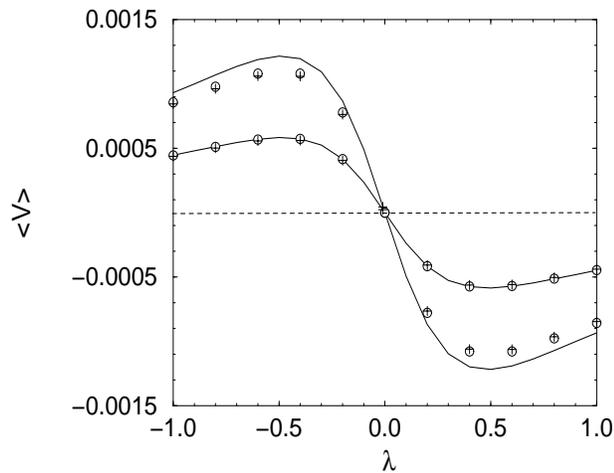}
\caption{\label{vel-lamb} Mean velocity of the kink center of mass
versus $\lambda$ for  $\epsilon=0.1$, $\omega=0.1$,
$\theta_{0}=\pi/2$ and for two values of the damping coefficient:
$\alpha=0.2$ and $\alpha=0.5$ (for fixed $\lambda$, the mean
velocity of the $\alpha=0.5$ case is smaller). }
\end{figure}

We have also investigated the dependence of the phenomenon on the
asymmetry parameter $\lambda$ as well as the importance of the
damping term for soliton ratchets. In Fig.\ \ref{vel-lamb} we show
the mean velocity as a function of $\lambda$ for fixed system
parameters and two different values of $\alpha$. We see that the
curve is anti-symmetric around the origin meaning that the sign of
$\lambda$ determines the direction of motion, the maximal effect
occurring around $|\lambda|=0.5$, i.e. the point of maximal
asymmetry of the potential.

\begin{figure}
\vspace*{2mm}
\includegraphics[width=6.5cm,height=8.0cm,angle=-90]{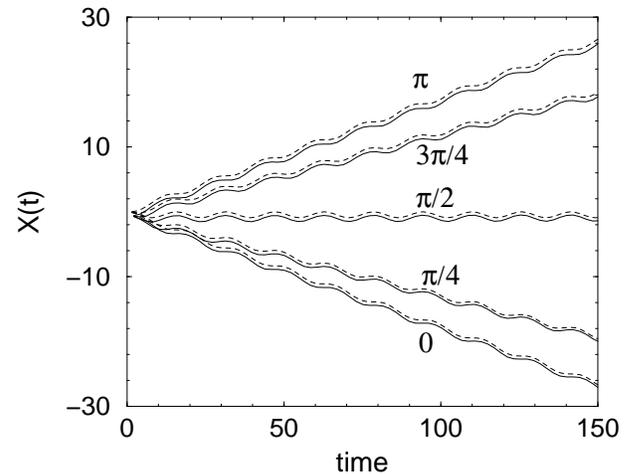}
\caption{\label{fig3} Time evolution of the center of the kink for
different values of initial phase, $\theta_{0}$, in the non damped
case. Parameters are fixed as: $\lambda=-0.5$; $\epsilon=0.1$ and
$\omega=0.4$. The solid and the dashed lines represent the
numerical simulations of Eq. (\ref{eq1}) and the numerical
solutions of Eq.s (\ref{cc1})-(\ref{cc3}), respectively.}
\end{figure}

Finally, we have investigated the zero damping limit of the
phenomenon. In Fig.\ref{fig3}  we depict the time evolution of the
center of mass of the kink for $\alpha=0$ and for different values
of the initial condition phase $\theta_{0}$ of the ac force. We
see that $X(t)$ is basically a linear function of time, whose
slope depends on $\cos(\theta_{0})$. This implies that, in this
case, is the initial phase which determines the direction of the
motion and not only the asymmetry of the potential, as predicted by the
CC analysis. Since in most experimental contexts initial phase are
usually unknown,  one must consider the phase as a random variable
and take average on it. This obviously implies that soliton
ratchets cannot exist in the zero-damping limit.

\section{IV. Conclusions}

In this paper  we have studied the ratchet dynamics of the kink
solution of the ADSGE by using a collective coordinate approach
with two collective variables, the center of mass and the width of
the kink. For these variables we have obtained a system of ODEs
from which we derived an approximated expression for the mean
velocity of the kink as a function of the system parameters. We
have confirmed that the CC approach is valid when phonons are not
excited in the system, i.e. for small values of $\epsilon$ and for
$\omega << \omega_{ph}$. We have shown that for a proper
description of soliton ratchets it is not enough to consider the
kink as a point particle moving  in a ratchet potential
\cite{goldobin,carapella}, but it is crucial to include also the
internal oscillations of the kink profile. In particular, we have
shown that the net motion of the kink becomes possible when its
internal and translational modes become {\it effectively} coupled
(the effective coupling being possible only in presence of
damping). We also showed that the asymmetry of the potential
determines the direction of the motion and that in the
zero-damping case the ratchet effect vanishes (i.e. it depends on
initial conditions). The resonant behavior of the velocity as a
function of frequency and damping was also investigated. We found
that the mean velocity approach a maximum value when the frequency
of the ac force goes to the internal mode frequency or when the
damping coefficient approach its optimal value. Finally, we have
shown that the occurrence of current reversal is related to the
presence of phonons in the system rather than the coupling between
the translational and the internal mode.

In conclusion, the results of our analysis confirm the internal
mode mechanism for soliton ratchet proposed in Ref. \cite{mn} and
provide an approximate analytical description of the phenomenon.

\begin{acknowledgments}
N.R.Q  acknowledge financial support from  the
Ministerio de Ciencia y Tecnolog\'{\i}a of Spain under the grant
BFM2001-3878-C02, and from the Junta de Andaluc\'{\i}a through the
project FQM-0207. M.S. acknowledges financial support
from the MURST, under a PRIN-2003 Initiative, and from the
European grant LOCNET, contract no. HPRN-CT-1999-00163.
\end{acknowledgments}

\end{document}